\documentclass[twocolumn,showpacs,pra]{revtex4}

\usepackage{amsmath,amssymb,amsfonts}

\renewcommand{\vec}[1]{\mathbf{#1}}

\begin{document}

\title{Photon polarization and Wigner's little group}

\author{Pawe{\l} Caban}\email{P.Caban@merlin.fic.uni.lodz.pl}
\author{Jakub Rembieli\'nski}\email{J.Rembielinski@merlin.fic.uni.lodz.pl}
\affiliation{Department of Theoretical Physics,
University of {\L}\'od\'z\\ 
Pomorska 149/153, 90-236 {\L}\'od\'z, Poland}

\date{\today}

\begin{abstract}
To discuss one-photon polarization states
we find the explicit form of the Wigner's little group element in the
massless case for arbitrary Lorentz transformation. As is well
known, when analyzing the transformation properties of the physical
states, only the value of the phase factor is relevant. We show that
this phase factor depends only on the direction of the momentum
$\vec{k}/|\vec{k}|$ and does not depend on the frequency $k^0$.
Finally, we use this observation to discuss the transformation
properties of the linearly polarized photons and the corresponding
reduced density matrix. We find that they transform properly under
Lorentz group.
\end{abstract}


\maketitle

\section{Introduction}

In recent years a lot of interest has been devoted to the study of the
quantum
entanglement and Einstein-Podolsky-Rosen correlation function under
the Lorentz transformations for massive particles 
\cite{cab_PST2002,cab_Czachor1997_1,cab_Czachor1997_2,cab_RS2002,cab_AM2002,%
cab_TU2002,cab_GA2002}. In recent papers
\cite{cab_LPT2003,cab_BGA2003} also the massless particle case was
discussed. One of the 
key ingredients of these papers is the calculation of the explicit
form of the little group element for massless particle in some special
cases to analyze transformation properties of entangled states and
reduced density matrix.

In this paper we derive the explicit form of the Wigner's little
group element in the massless case for an arbitrary Lorentz
transformation and  discuss the transformation
properties of the linearly polarized photons and the corresponding
reduced density matrix obtained by tracing out kinematical degrees of
freedom. As is well known in the Hilbert space of massless particles
the one-particle momentum eigenvectors under Lorentz transformation
$\Lambda$ are multiplied by a phase factor depending on $\Lambda$ and
the particle four-momentum $k^\mu$. We show that
this phase factor depends only on $\Lambda$ and the direction of the
momentum $\vec{k}/|\vec{k}|$ but does not depend on the frequency
$k^0$. In contrast to other papers \cite{cab_LPT2003} this observation
enables us to give the description of the transformation rules of the
linearly polarized photons which are not necessarily monochromatic.

\section{Wigner's little group for massless particles}

As is well known, the pure quantum states are identified with rays in
the Hilbert space. For this reason, on the quantum level, we should use
ray representations of the classical symmetry groups. In our case of
the proper ortochronous Poincar\'e group $P_{+}^{\uparrow}$, which is
the semidirect product of the proper ortochronous
Lorentz group $L_{+}^{\uparrow}$ and the translations group $T^4$, its
ray representations (so called double-valued representations) are
faithful representations of the universal 
covering of $P_{+}^{\uparrow}$, i.e., the semidirect product of
$SL(2,{\mathbb{C}})$ and $T^4$. Moreover, the faithful
representations of $P_{+}^{\uparrow}$ are homomorphic representations
of its universal covering group.

We use the canonical homomorphism between the group
$SL(2,{\mathbb{C}})$ (universal covering of the proper ortochronous
Lorentz group $L_{+}^{\uparrow}$) and the Lorentz group
$L_{+}^{\uparrow}\sim SO(1,3)_0$ \cite{cab_BR1977}.  
This homomorphism is defined as follows:
With every four-vector $k^\mu$ we associate a two-dimensional
Hermitian matrix ${\textsf k}$ such that
 \begin{equation}
 {\textsf k}=k^\mu \sigma_\mu,
 \label{matrix_k}
 \end{equation}
where $\sigma_i,i=1,2,3$ are the standard Pauli matrices
and $\sigma_0=I$. In the space of two-dimensional Hermitian matrices
(\ref{matrix_k}) the Lorentz group action is given by 
 \begin{equation}
 {\sf k}^\prime=A{\textsf k}A^\dag,
 \label{lorentz_group_action}
 \end{equation}
where $A$ denotes element of the
$SL(2,{\mathbb{C}})$ group corresponding to the Lorentz transformation 
$\Lambda(A)$ which converts the four-vector $k$ to $k^\prime$ (i.e.,
${k^\prime}^\mu=\Lambda_{\phantom{\mu}\nu}^{\mu}k^\nu$) and
${\textsf k}^\prime={k^\prime}^\mu\sigma_\mu$. The kernel of
this homomorphism is isomorphic to ${\mathbb{Z}}_2$ (the center of the
$SL(2,{\mathbb{C}})$).

Now, let us focus on the case of massless particles. 
An explicit matrix representation (\ref{matrix_k}) of the null
(light-cone) four-vector $k$ can be written as
 \begin{equation}
 {\textsf k}=k^0\begin{pmatrix}
 1+n^3 & n_- \\ n_+ & 1-n^3
 \end{pmatrix},
 \label{matrix_k_explicit}
 \end{equation}
where $n_\pm=n^1\pm in^2$,
$\vec{n}=\vec{k}/|\vec{k}|$, $k^0=|\vec{k}|$ and $\det{\textsf k}=k^\mu
k_\mu=0$.  
In this case we choose the standard four-vector as
$\tilde{k}=(1,0,0,1)$. In the matrix representation
(\ref{matrix_k_explicit}) the following matrix is associated with
$\tilde{k}$:
 \begin{equation}
 \tilde{\textsf k}=\begin{pmatrix}
 2 & 0 \\ 0 & 0
 \end{pmatrix}.
 \end{equation}
Now, let us find the stability group of $\tilde{{\textsf k}}$, i.e., 
$A_0\in SL(2,{\mathbb{C}})$ which
leaves $\tilde{\textsf k}$ invariant. All such $A_0$ form a
subgroup of the $SL(2,{\mathbb{C}})$ group, i.e. 
 \begin{equation}
 (\text{stability group})=\{A_0\in SL(2,{\mathbb{C}})\colon
 \tilde{\textsf k}=A_0{\tilde{\textsf k}}A_{0}^{\dag} \}.
 \end{equation}
As is well known \cite{cab_BR1977} the stability group of
the four-vector 
$\tilde{k}$ is isomorphic to 
the $E(2)$ group of rigid motions of Euclidean plane.
We can easily find the most general $A_0$ by
solving the equation
$\tilde{\textsf k}=A_0{\tilde{\textsf k}}A_{0}^{\dag}$. We get 
 \begin{equation}
 A_0=\begin{pmatrix}
 e^{i\frac{\psi}{2}} & z \\ 0 & e^{-i\frac{\psi}{2}}
 \label{little_group_general}
 \end{pmatrix},
 \end{equation}
where $z$ is an arbitrary
complex number. Since the $SL(2,{\mathbb{C}})$ is the two-fold covering
of the Lorentz group, we restricted the values of $\psi$ to the interval
$\left\langle0,2\pi\right)$.
Our purpose is to find the Wigner's little group element $W(\Lambda,k)$
corresponding to $k$ and the Lorentz transformation 
$\Lambda$, namely
 \begin{equation}
 W(\Lambda,k)=L^{-1}_{\Lambda k}\Lambda L_k,
 \end{equation}
where $L_k\in L_{+}^{\uparrow}$ is determined uniquely by the
following conditions:
 \begin{equation}
 k=L_k\tilde{k},\quad L_{\tilde{k}}=I.
 \label{conditions_L_k}
 \end{equation}
In order to find the corresponding element $S(\Lambda,k)$ in
$SL(2,{\mathbb{C}})$ such that $W(\Lambda,k)=\Lambda(S(\Lambda,k))$,
i.e.\ 
 \begin{equation}
 S(\Lambda,k)=A^{-1}_{\Lambda k} A A_k, 
 \label{little_group_two_dimensional}
 \end{equation}
where $\Lambda(A_k)=L_k$,
we have to first calculate the matrix $A_k$. We can do it by solving
the matrix equation
 \begin{equation}
 {\textsf k}=A_k\tilde{\textsf k}A^{\dag}_{k}.
 \end{equation}
After simple calculation we get
 \begin{equation}
 A_k=U_{\vec{n}}B(k^0),
 \end{equation}
where
 \begin{equation}
 U_{\vec{n}}=\frac{1}{\sqrt{2(1+n^3)}}\begin{pmatrix}
 1+n^3 & -n_- \\ n_+ & 1+n^3
 \end{pmatrix}
 \end{equation}
represents rotation $R_{\vec{n}}$ which converts the spatial vector
$(0,0,1)$ to $\vec{n}$, while
 \begin{equation}
 B(k^0)=\begin{pmatrix}
 \sqrt{k^0} & 0 \\ 0 & \frac{1}{\sqrt{k^0}}
 \end{pmatrix}
 \end{equation}
represents boost along $z$-axis which converts $\tilde{k}$ to
$k^0\tilde{k}$. Therefore
 \begin{equation}
 A_k=\frac{1}{\sqrt{2k^0(1+n^3)}}\begin{pmatrix}
 k^0(1+n^3) & -n_- \\ k^0n_+ & 1+ n^3
 \end{pmatrix}.
 \label{A_k}
 \end{equation}
Note that according to Eq.\ (\ref{conditions_L_k}) $A_{\tilde{k}}=I$.
Now, an arbitrary Lorentz transformation $\Lambda(A)$ is represented
in $SL(2,{\mathbb{C}})$ by the corresponding complex unimodular matrix
 \begin{equation}
 A=\begin{pmatrix}
 \alpha & \beta \\ \gamma & \delta
 \end{pmatrix},\quad \alpha\delta-\beta\gamma=1.
 \end{equation}
To calculate $A_{\Lambda k}$ we simply use the formulas
(\ref{lorentz_group_action},\ref{matrix_k_explicit}) to find
${\textsf k}^\prime$  and then identify $k^\prime=\Lambda k$. We find
 \begin{align}
 {k^\prime}^0  = & \frac{1}{2}k^0 {a} \label{k_0}\\
 {n^\prime}^3  = &\frac{2{b}}{{a}} -1,\label{n_3}\\
 n^{\prime}_{+}  = &\frac{2c}{{a}},\\
 n^{\prime}_{-} = & {n^{\prime}_{+}}^*,\label{n_-}
 \end{align}
where
 \begin{align}
 {a} = & (|\alpha|^2+|\gamma|^2)(1+n^3)+
 (|\beta|^2+|\delta|^2)(1-n^3)\nonumber\\
 & +(\alpha\beta^*+\gamma\delta^*)n_- +
 (\alpha^*\beta+\gamma^*\delta)n_+,\label{cab_a}\\ 
 {b} = & |\alpha|^2(1+n^3)+|\beta|^2(1-n^3)+\alpha\beta^*n_- 
  +\alpha^*\beta n_+,\label{cab_b}\\
 {c} = & \alpha^*\gamma(1+n^3) +
 \beta^*\delta(1-n^3) + \beta^*\gamma n_- + \alpha^*\delta n_+,
 \label{cab_c} 
 \end{align}
and
${\vec{n}}^\prime={\vec{k}}^\prime/|{\vec{k}}^\prime|$. 
Therefore we can find the explicit form of $S(\Lambda,k)$ by means
of Eqs.\ (\ref{little_group_two_dimensional}) and (\ref{A_k}). 
We have to calculate only the
elements $S(\Lambda,k)_{11}$ and $S(\Lambda,k)_{12}$,
since the general little group 
element (\ref{little_group_general}) depends only on the phase factor
$e^{i\frac{\psi}{2}}$ and complex number $z$.
A straightforward calculation yields finally the following formulas:
 \begin{align}
 e^{i\frac{\psi(\Lambda,k)}{2}} = & 
 \frac{(\alpha(1+n^3)+\beta n_+){b}+(\gamma(1+n^3)+\delta n_+){c}^*}%
 {{a}\sqrt{{b}(1+n^3)}} 
 \label{phase}\\
 z(\Lambda,k) = & \frac{(-\alpha n_-+\beta(1+n^3)){b} + (-\gamma
 n_-+\delta(1+n^3)){c}^*}{k^0 {a} \sqrt{{b}(1+n^3)}}  \label{cab_z}
 \end{align}
where $a,b,$ and $c$ are given by Eqs.\ (\ref{cab_a}-\ref{cab_c}).

The unitary
irreducible representations of the Poincar\'e group are induced from
the unitary irreducible representations of the little group of the
four-momentum $k^\mu$ (i.e., the $E(2)$ group in the case of the
massless particles) \cite{cab_Weinberg1964,cab_BR1977} . Now, we have
two classes of the unitary irreducible representations of 
$E(2)$: the faithful infinite dimensional
representations and the one-dimensional homomorphic representations of
$E(2)$, isomorphic to its compact subgroup $SO(2)\subset
E(2)$. Because there is no evidence for existence of massless particles
with a continuous 
intrinsic degrees of freedom the physical choice is the last one
\cite{cab_Weinberg1964}. Thus 
by means of the induction procedure \cite{cab_BR1977} the four-momentum
eigenstates transform according to the formula
 \begin{equation}
 U(\Lambda)\left|k,\lambda\right\rangle = e^{i\lambda\psi(\Lambda,k)}
 \left|\Lambda k,\lambda\right\rangle.
 \end{equation}
In the above formula $U(\Lambda)$ denotes unitary operator
representing $\Lambda$ in the unitary representation of the Poincar\'e
group while the helicity $\lambda$ fixes irreducible unitary
representation of the Poincar\'e group induced from $SO(2)$; $\lambda$
takes integer and half-integer values only
\cite{cab_Weinberg1964,cab_BR1977}. 
We use invariant normalization of the four-momentum eigenstates
$\left|k,\lambda\right\rangle$, i.e., $\left\langle
  p,\sigma|k,\lambda\right\rangle = 2 k^0
\delta_{\sigma\lambda}\delta(\vec{k}-\vec{p})$. Thus, when analyzing
the transformation properties of physical states only the value of the
phase $\psi(\Lambda,k)$ is relevant (Eq.\ (\ref{phase})). So it is
very important to stress that {\em 
the value of the phase $\psi$ depends only on $\Lambda$ and $\vec{n}$
and does not depend on the frequency $k^0$}:
 \begin{equation}
 \psi(\Lambda,k)=\psi(\Lambda,\frac{\vec{k}}{|\vec{k}|}) =
 \psi(\Lambda,\vec{n}). 
 \label{phase_direction}
 \end{equation}

Note also that momenta of massless particles which are parallel
in one inertial frame are parallel for every inertial
observer, i.e.,
 \begin{equation}
 \frac{\vec{k}}{|\vec{k}|} = \frac{\vec{p}}{|\vec{p}|}
 \Rightarrow  
 \frac{\vec{k}^{\prime}}{|\vec{k}^{\prime}|} =
 \frac{\vec{p}^{\prime}}{|\vec{p}^{\prime}|},
 \label{paralel}
 \end{equation}
where $k^{\prime} = \Lambda k$, $p^{\prime}=\Lambda p$. Indeed, for
massless particles, 
$\vec{k}$ and $\vec{p}$ are parallel iff the corresponding four-momenta
are Lorentz orthogonal, i.e., $k_\mu p^\mu=0$. Since
$k_\mu p^\mu$ is a Lorentz invariant then this holds in all inertial
frames. Equation (\ref{paralel}) can be also verified explicitly by
using  Eqs.\ (\ref{n_3}-\ref{n_-}).
The above property holds good only in the massless case. 

Now, using Eq.\ (\ref{phase}) we can immediately obtain the value of
$e^{i\psi(\Lambda,k)}$ in a number of special cases considered
elsewhere. 

{\textbf{Rotations:}} In the case $\Lambda=R$ we have
$R=\Lambda(U)$ where $U\in SU(2)\subset SL(2,{\mathbb{C}})$, thus 
we put
 \begin{equation}
 \delta=\alpha^*,\quad \gamma=-\beta^*,\quad |\alpha|^2+|\beta|^2=1 
 \end{equation}
and from (\ref{phase}) we get the following simple formula:
 \begin{equation}
 e^{i\psi(R,k)}=\frac{\alpha(1+n^3)+\beta
 n_+}{\alpha^*(1+n^3)+\beta^* n_-}.
 \label{phase_rotation}
 \end{equation}
For the given rotation $R$ the explicit form of $\alpha$ and $\beta$
can be expressed by, e.g., Euler angles (see, for example, Ref.\
\cite{cab_KK1961}). Also note that from Eq.\ (\ref{cab_z}), we
get  
 \begin{equation}
 z(R,k)=0.
 \end{equation}
Now let us consider the special case of the rotation
$R(\chi\vec{n})$ around the direction ${\vec{n}}$. 
The matrix $U_{R(\chi\vec{n})}\in SU(2)$ representing
$R(\chi\vec{n})$ can be written in the form \cite{cab_Gursey1963} 
 \begin{equation}
 U_{R(\chi\vec{n})}=e^{\frac{i}{2}\chi n^j \sigma_j} =
 \cos{\frac{\chi}{2}}+in^j\sigma_j 
 \sin{\frac{\chi}{2}}.
 \end{equation}
Thus,
 \begin{equation}
 U_{R(\chi\vec{n})}=\begin{pmatrix}
 \cos{\frac{\chi}{2}}+in^3\sin{\frac{\chi}{2}} & in_-\sin{\frac{\chi}{2}}\\
 in_+\sin{\frac{\chi}{2}} & \cos{\frac{\chi}{2}}-in^3\sin{\frac{\chi}{2}}
 \end{pmatrix}.
 \end{equation}
Thus inserting the corresponding values of $\alpha$ and
$\beta$ to Eq.\ (\ref{phase_rotation}), we find in this case 
 \begin{equation}
 e^{i\psi(R(\chi\vec{n}),\vec{n})}=e^{i\chi}.
 \end{equation}
As the next example, we consider the rotation
$R(\chi\hat{\vec{z}})$ around the $z$-axis. In this case (see Ref.\ 
\cite{cab_Gursey1963}) 
 \begin{equation}
 U_{R(\chi\hat{\vec{z}})}= \begin{pmatrix}
 e^{i\frac{\chi}{2}} & 0 \\ 0 & e^{-i\frac{\chi}{2}}
 \end{pmatrix},
 \end{equation}
therefore from Eq.\ (\ref{phase_rotation}), we get the same formula
as previously 
 \begin{equation}
 e^{i\psi(R(\chi\hat{\vec{z}}),\vec{n})}=e^{i\chi}.
 \end{equation}

{\textbf{Boosts:}} Pure Lorentz boost $\Lambda(\vec{v})$ in an
arbitrary direction 
$\vec{e}=\frac{\vec{v}}{|\vec{v}|}$ can be represented by
the following 
$SL(2,{\mathbb{C}})$ matrix \cite{cab_BR1977}:
 \begin{equation}
 A(\vec{v})=e^{\frac{1}{2}\xi e^j\sigma_j}=
 \begin{pmatrix}
 \cosh{\frac{\xi}{2}} + e^3 \sinh{\frac{\xi}{2}} &
 e_- \sinh{\frac{\xi}{2}}  \\
 e_+ \sinh{\frac{\xi}{2}} & \cosh{\frac{\xi}{2}} - e^3
 \sinh{\frac{\xi}{2}} 
 \end{pmatrix},
 \end{equation}
where the parameter $\xi$ is connected with the velocity of the
boosted frame by the relation
 \begin{equation}
 \tanh{\xi}=-v
 \end{equation}
and $e_{\pm}=e^1\pm i e^2$ 
(we use natural units with the light velocity equal to 1).
Inserting the corresponding values of $\alpha$ and $\beta$ into 
Eqs.\ (\ref{cab_a}-\ref{cab_c}), we arrive at the relations of the form 
 \begin{align}
 a & = 2(\cosh{\xi}+\vec{e}\cdot\vec{n}\sinh{\xi}), \label{cab_a_boost} \\ 
 b & = \cosh{\xi} + n^3 + (e^3+\vec{e}\cdot\vec{n})\sinh{\xi} + e^3
 \vec{e}\cdot\vec{n} (\cosh{\xi}-1), \label{cab_b_boost}\\
 c & = n_+ +(\sinh{\xi}+\vec{e}\cdot\vec{n}(\cosh{\xi}-1))e_+.
 \label{cab_c_boost}
 \end{align}
Now, the corresponding little group element can be obtained from Eqs.\
(\ref{phase},\ref{cab_z}). We consider here two special cases: boost
$A(v\vec{n})$ 
along the $\vec{n}$ direction and boost $\Lambda(v\hat{\vec{z}})$
along the $z$ direction. In the
first case, we have $\vec{e}=\vec{n}$ and from Eqs.\
(\ref{cab_a_boost}-\ref{cab_c_boost}) and (\ref{phase},\ref{cab_z}), we
find 
 \begin{equation}
 e^{i\psi(\Lambda(v\vec{n}),k)}=1, \quad z(\Lambda(v\vec{n}),k)=0.
 \end{equation}
In the second case, we have 
 \begin{equation}
 \alpha=\frac{1}{\delta}=\sqrt[4]{\frac{1-v}{1+v}},\quad  \beta=\gamma=0.
 \end{equation}
Inserting above values to Eqs.\ (\ref{phase}) and (\ref{cab_z}), we get
 \begin{equation}
 \psi(\Lambda(v\hat{\vec{z}}),k)=0, \quad
 z(\Lambda(v\hat{\vec{z}}),k)=\frac{n_-}{k^0(\frac{1}{v}-n^3)}.
 \end{equation}

\section{Transformation law for linearly polarized light}

Now, we apply the above results to discuss some transformation
properties of the polarized states and reduced density matrix for
photons. 

Let us consider first the classical electromagnetic field. As is
well known the monochromatic plane electromagnetic wave is in general
polarized eliptically. In the special case of the linear polarization
we can deal also
with the plane wave which is not necessarily monochromatic. In this
case the electromagnetic field tensor can be written as
 \begin{equation}
 F^{\mu\nu}(x)={\mathcal{F}}^{\mu\nu}f(ct-\vec{n}\cdot\vec{x})
 \label{linear_clasical}
 \end{equation}
for the lineary polarized wave propagating in the $\vec{n}$
direction, where the tensor ${\mathcal{F}}^{\mu\nu}$ is $x^\mu$
independent. It is evident that the above formula is covariant under
Lorentz transformations. It means that Lorentz 
transformations preserve linear polarization of an arbitrary plane
wave (not necessarily monochromatic). We show that it is also the case
on the quantum level. 

As is well known (see, e.g., Refs.\ 
[\onlinecite{cab_Weinberg1964},\onlinecite{cab_Weinberg1996}]) the
one-photon representation space is 
spanned by the vectors $\{\left|k,1\right\rangle,
\left|k,-1\right\rangle\}$ because the parity operator changes the
sign of the helicity. Let us consider first the
monochromatic linearly polarized plane wave. 
The photon state corresponding to such a wave is of the form
\cite{cab_Weinberg1996}
 \begin{equation}
 \left|k,\phi\right\rangle \equiv
 \left|(k^0,|\vec{k}|\vec{n}),\phi\right\rangle = \frac{1}{\sqrt{2}}
 \sum_{\substack{\lambda=-1 \\ \lambda\not=0}}^{1} e^{i\lambda\phi}
 \left|k,\lambda\right\rangle,  
 \label{monochromatic_photon}
 \end{equation}
where $k^0=|\vec{k}|$ and the momentum independent angle $\phi$
determines the direction of 
the polarization in the plane perpendicular to the direction of the
propagation $\vec{n}$. The general linearly polarized state
corresponding to the wave (\ref{linear_clasical}) has the form
 \begin{equation}
 \left|g,\phi,\vec{n}\right\rangle = \frac{1}{\sqrt{2}}
 \sum_{\lambda} e^{i\lambda\phi} \int_{0}^{\infty} d|\vec{k}| \,
 g(|\vec{k}|)\, \left|(k^0,|\vec{k}|\vec{n}),\lambda\right\rangle,
 \label{general_lineary_polarized}
 \end{equation}
where $\vec{n}$ is fixed. The state
(\ref{general_lineary_polarized}) is a tensor product of momentum
direction and polarization states in each Lorentz frame.
Let us note that states belonging to the
proper Hilbert space (wave packets) cannot be exactly linearly
polarized states. However, linearly polarized states
(\ref{general_lineary_polarized}) can be approximated (as tempered
distributions) with an arbitrary accuracy by sequences of wave
packets. It is interesting to point out a parallelism between classical
and quantum description of ideal linearly polarized states. Namely, on
the classical level they have infinite total electromagnetic energy
while on the quantum level they lie out of the proper Hilbert space of
the wave packets, i.e., they are distributions.

Now, we show that {\em for every inertial observer the lineary
polarized state 
(\ref{general_lineary_polarized}) remains lineary polarized.} Indeed,
taking into account Eqs.\ (\ref{phase}-\ref{phase_direction}) we
find
 \begin{align}
 U(\Lambda) \left|g,\phi,\vec{n}\right\rangle =&
 \frac{1}{\sqrt{2}}
 \sum_{\lambda} e^{i\lambda(\phi+\psi(\Lambda,\vec{n}))} \nonumber\\
 &  \times\int_{0}^{\infty} d|\vec{k}| \,
 g^\prime(|\vec{k}|)\,
 |({k}^0,|\vec{k}|\vec{n}^\prime),\lambda\rangle \nonumber\\
 = & \left|g^\prime,\phi+\psi,{\vec{n}}^\prime \right\rangle,
 \label{transformation_linery_polarized}
 \end{align}
where 
 \begin{equation}
 g^\prime(|\vec{k}|) = \frac{2}{a}
 g(\frac{2|\vec{k}|}{a}),
 \end{equation}
$a$ is given by Eq.\ (\ref{cab_a}) and by virtue of Eq.\
(\ref{paralel}) the direction $\vec{n}^\prime$ is fixed,
too. Therefore, the state we received is again linearly 
polarized.

Now, we discuss the transformation of the reduced density matrix for
lineary polarized plane wave. In general for the reduced density
matrix describing the helicity properties of the state
 \begin{equation}
 |f\rangle = \sum_{\lambda}\int d\mu(k)f_\lambda(k)|k,\lambda\rangle 
 \end{equation}
we obtain the following formula:
 \begin{equation}
 \hat{\rho}_{\sigma\lambda} =
 \frac{\int d\mu(k)f_\sigma(k)f_{\lambda}^{*}(k)}%
 {\sum_\lambda\int d\mu(k)|f_\lambda(k)|^2},
 \label{reduced_general}
 \end{equation}
where we have used the Lorentz invariant measure
 \begin{equation}
 d\mu(k)=\theta(k^0)\delta(k^2)\, d^4k \equiv
 \frac{d^3\vec{k}}{2|\vec{k}|}.
 \end{equation}
It should be noted that in general the state $|f\rangle$ can be a tempered
distribution (it does not necessarily
belong to the Hilbert space but rather to the Gel'fand
triple), as for example four-momentum
eigenstates. In such a situation the formula 
(\ref{reduced_general}) should be understood as a result of a proper
regularization procedure. Applying the above considerations to the density
matrix describing the state $\left|g,\phi,\vec{n}\right\rangle$ we get
the following reduced density matrix:
 \begin{equation}
 \rho_{\lambda\sigma}(g,\phi,\vec{n}) =
 \frac{1}{2}e^{i(\lambda-\sigma)\phi}, 
 \end{equation}
i.e.,
 \begin{equation}
 \rho(g,\phi,\vec{n}) = \frac{1}{2}\begin{pmatrix}
 1 & e^{2i\phi} \\ e^{-2i\phi} & 1
 \end{pmatrix},\label{reduced_pure}
 \end{equation}
which in fact represents a reduced pure state because $\rho^2=\rho$. 
The above density matrix transforms properly under Lorentz
transformations, namely 
 \begin{align}
 \rho^\prime = & 
 \begin{pmatrix}
 e^{i\psi(\Lambda,\vec{n})} & 0 \\ 0 & e^{-i\psi(\Lambda,\vec{n})} 
 \end{pmatrix}
 \rho(g,\phi,\vec{n})
 \begin{pmatrix}
 e^{-i\psi(\Lambda,\vec{n})} & 0 \\ 0 & e^{i\psi(\Lambda,\vec{n})} 
 \end{pmatrix} \nonumber\\
 = & \frac{1}{2}
 \begin{pmatrix}
 1 & e^{2i(\phi+\psi)} \\ e^{-2i(\phi+\psi)} & 1
 \end{pmatrix} =
 \rho(g^\prime,\phi+\psi,\vec{n}^\prime).\label{transformation_density} 
 \end{align}
We stress that the fact that the linearly polarized state admits a
covariant reduced density matrix description in terms of helicity
degrees of freedom is related to the property of the Lorentz
transformation that it does not generate entanglement between momentum 
direction and helicity.

Finally, let us note that the von Neumann entropy
corresponding to the density matrix (\ref{reduced_pure}) is equal to 
zero. Evidently it is Lorentz-invariant in view of Eq.\
(\ref{transformation_density}). 

Our discussion can be easily recast in terms of
polarization vectors defined according to Ref.\
\cite{cab_Weinberg1964}, for different approach see also Ref.\
\cite{cab_PT2003_2}. 
 
\section{Conclusions}

We have found in this paper the explicit form of the Wigner's little group
element in the massless case for arbitrary Lorentz
transformation. Using this result we have shown that the light wave
which is linearly polarized (but not necessarily monochromatic) for one
inertial observer remains linearly polarized also for an arbitrary inertial
observer. We have also shown that the reduced density matrix
describing linearly polarized photon, obtained
by tracing out kinematical degrees of freedom, transforms properly
under Lorentz group 
action. Moreover the corresponding von Neumann entropy is a Lorentz
scalar.

\begin{acknowledgments}
This work was supported by University of {\L}\'od\'z and the
Laboratory of Physical Bases of Processing of Information.
\end{acknowledgments}


\end{document}